\documentclass{article}

\PassOptionsToPackage{numbers, compress}{natbib}

\usepackage[preprint]{neurips_2019}

\usepackage[utf8]{inputenc} 
\usepackage[T1]{fontenc}    
\usepackage{hyperref}       
\usepackage{url}            
\usepackage{booktabs}       
\usepackage{amsfonts}       
\usepackage{nicefrac}       
\usepackage{microtype}      

\usepackage[dvipdfmx]{graphicx}

\usepackage{amsmath} 
\usepackage{listings}
\usepackage{color}

\title{Computational cost for determining an approximate global minimum using the selection and crossover algorithm}

\author{
  Takuya Isomura \\
  RIKEN Center for Brain Science \\
  Wako, Saitama 351-0198, Japan \\
  \texttt{takuya.isomura@riken.jp} \\
}

\date{}

\def \E {\mathrm{E}}
\def \Var {\mathrm{Var}}

\begin{document}

\maketitle

\begin{abstract}
This work examines the expected computational cost to determine an approximate global minimum of a class of cost functions characterized by the variance of coefficients. The cost function takes $N$-dimensional binary states as arguments and has many local minima. Iterations in the order of $2^N$ are required to determine an approximate global minimum using random search. This work analytically and numerically demonstrates that the selection and crossover algorithm with random initialization can reduce the required computational cost (i.e., number of iterations) for identifying an approximate global minimum to the order of $\lambda^N$ with $\lambda$ less than 2. The two best solutions, referred to as parents, are selected from a pool of randomly sampled states. Offspring generated by crossovers of the parents' states are distributed with a mean cost lower than that of the original distribution that generated the parents. It is revealed that in contrast to the mean, the variance of the cost of the offspring is asymptotically the same as that of the original distribution. Consequently, sampling from the offspring's distribution leads to a higher chance of determining an approximate global minimum than sampling from the original distribution, thereby accelerating the global search. This feature is distinct from the distribution obtained by a mixture of a large population of favorable states, which leads to a lower variance of offspring. These findings demonstrate the advantage of the crossover between two favorable states over a mixture of many favorable states for an efficient determination of an approximate global minimum.
\end{abstract}


\section{Introduction}

Biological organisms occasionally outperform current machine learning techniques in terms of adaptive ability and robustness. This work examines the benefits of the search and optimization strategy used by organisms, selection and crossover \cite{Mitchell1998, Sivanandam_Deepa2007, Weise2009, Lumley2015}, toward the creation of a biologically-inspired optimization framework.

The intelligence of organisms refers to the optimization of search algorithms to identify optimal solutions. Organisms can recognize their surrounding environment by optimizing their internal representations about the dynamics and cases in the external world. In addition, they can optimize their behavior for adapting to the environment to increase the probability of survival and reproduction. These optimization problems are in many cases formulated as the minimization of a cost function. Theoretical neurobiology commonly uses cost functions to model neural activity \cite{Bourdoukan2012, Boerlin2013}, various learning processes \cite{Linsker1988, Dayan1995, Sutton_Barto1998, Brown2001, Knill_Pouget2004, Friston2006}, and evolution of nervous system \cite{Friston2010}. A common optimization strategy under a cost function is gradient descent, a form of local search, that updates neural activities, synaptic strengths, and genes, to minimize the cost function. However, the gradient descent is not effective when the cost function has a huge number of local minima as a state is likely to be trapped in a bad local minimum.

Apart from local search algorithms, the selection and crossover algorithm, which is a nonlocal search method, has been popularized in the literature on genetic algorithm \cite{Mitchell1998, Sivanandam_Deepa2007, Weise2009}. Crossovers can generate offspring that are sufficiently different from their parents, while inheriting some aspects of their favorable properties, when the parents' states are sufficiently different from each other. This can provide a search strategy that is different from the local search algorithms, and may thus provide a higher chance of determining an approximate global minimum. Although the benefits of selection and crossover have been proposed as the building block hypothesis \cite{Forrest_Mitchell1993, Stephens_Waelbroeck1999}, however, the conditions under which the selection and crossover algorithm accelerates the search remain unclear.

This work analytically and numerically explores the computational cost of the selection and crossover algorithm to determine an approximate global minimum, under the assumption that the output value of the cost function (i.e., its histogram) follows a Gaussian distribution (Fig. 1A). It is analytically expressed that problems requiring the $2^N$-order time using a simple random search can be solved in the $\lambda^N$-order time, with $1 < \lambda < 2$, using the selection and crossover algorithm. This finding highlights the utility of selection and crossover for accelerating the global search.


\section{Methods}
\subsection{System and cost function}

Here, we consider an optimization problem of $N$-dimensional binary states. A state is expressed using vector $x \equiv (x_1, \dots, x_N) \in X$ whose elements takes $-1$ or $1$, where $X \equiv \{ -1, 1 \}^N$. A cost function of this system $F(x): X \mapsto \mathbb{R}$ maps $x$ to a real number that expresses the cost of position $x$ in $N$-dimensional space. The aim of this work is to identify $x$ that provides an approximate global minimum of $F(x)$. In general, $F(x)$ is parameterized by up to $2^N$ independent parameters; this is a general setup for combinational optimization problems \cite{Korte_Vygen2012}. Variables and parameter definitions are provided in Supplementary Table S1. In relation to neurobiology, $x$ can be assumed to represent neural activity, synaptic connections, or genes; thus, this optimization problem can become a model of perception, learning, or evolution, respectively.

We express $F(x)$ as the sum of products. Because $x_i$ takes only $-1$ or $1$, $F(x)$ is expanded as
\begin{align}
\label{cost_function}
F(x) = \sum_i a_i x_i + \sum_{i < j} a_{ij} x_i x_j + \sum_{i < j < k} \!\! a_{ijk} x_i x_j x_k + \cdots + \sum_i a_{1 \dots N \backslash i} \prod_{j \neq i} x_j + a_{1 \dots N} \prod_i x_i \nonumber \\
= \sum_{1 \leq \alpha \leq N} \left( \sum_{1 \leq i_1 < \cdots < i_\alpha \leq N} \!\!\!\!\!\!\!\!\!\! a_{i_1 \dots i_\alpha} x_{i_1} \cdots x_{i_\alpha} \right). \hspace{40mm}
\end{align}
Here, coefficients $\{ a_i, a_{ij}, a_{ijk}, \dots, a_{1 \dots N \backslash i}, a_{1 \dots N} \}$, or $a_{i_1 \dots i_\alpha}$ in general using indices $i_1,\dots,i_\alpha$, are assumed to independently follow Gaussian distributions:
\begin{equation}
a_{i_1 \dots i_\alpha} \sim \mathcal{N}[\mu_\alpha, \sigma_\alpha^2]
\end{equation}
for $1 \leq \alpha \leq N$. Without loss of generality, it is assumed that $\mu_1 = \cdots = \mu_N = 0$, so that $\E_X[F(x)] = 0$, and $\Var_X[F(x)] = \E_X[F(x)^2] = 1$, as the cost function can be rescaled prior to optimization. Here, $\E_X[\bullet(x)]$ refers to the expectation of $\bullet(x)$ when elements of $x$ are independently and randomly selected from the distribution over $X$. Let $x_i$ ($1 \leq i \leq N$) take 1 with a probability of $1/2$. Immediately, as the expectation over various $a_{i_1 \dots i_\alpha} \in A$, where $A$ indicates the distribution of coefficients, it holds that
\begin{equation}
\label{variance}
\E_A\Big[ \Var_{X}[F(x)] \Big] = \sum_{1 \leq \alpha \leq N} \binom{N}{\alpha}\sigma_\alpha^2 = 1
\end{equation}
using the binomial coefficient $\binom{N}{\alpha} \equiv N! / (\alpha! (N-\alpha)!)$. In addition, it is assumed that $F(x)$ and $F(y)$ are nearly independent of each other when $x \neq y$. When $N$ is large, the distribution of $F(x)$ asymptotically approaches the unit Gaussian distribution, $p(F) = \mathcal{N}[0, 1]$, according to the central limit theorem (Fig. 1A). Under these assumptions, in what follows, we will estimate the distribution of an approximate global minimum and the required computational cost to determine it.


\begin{figure}[t]
  \centering
  \includegraphics[width=0.98\linewidth]{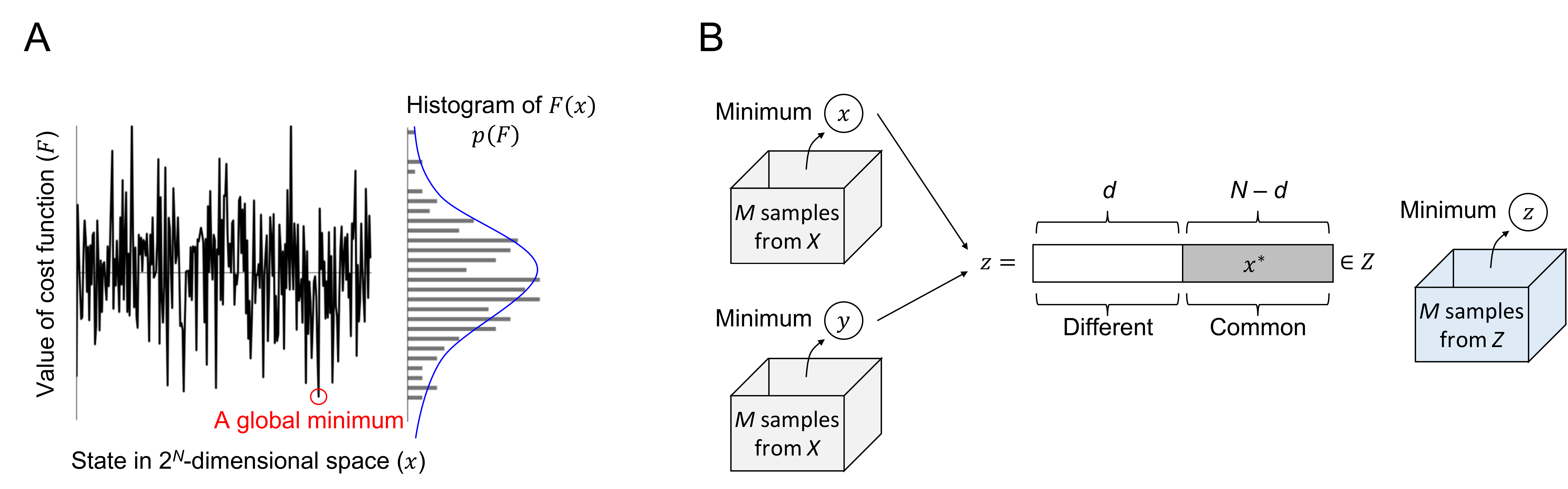}
  \caption{Schematics of system and procedure. (A) Schematic of cost function. The output value of the cost function follows a unit Gaussian distribution. (B) Schematic of selection and crossover. A crossover of two states $x,y$ that provide the minima generates offspring $z$.}
\end{figure}


\subsection{Approximate global minimum and computational cost of random search}

The optimization problem is simplified if only an approximate global minimum, not the exact global minimum, must be identified. The minimum among randomly selected $M$ samples from a set $\Omega \subseteq X$ can be estimated when their costs follow a Gaussian distribution. Thus, an approximate global minimum of our system can be defined as the minimum among randomly selected $2^N$ samples from $X$. Using the Laplace approximation, the probability distribution of the minimum of $M$-times random search is obtained as follows.

Let us independently and randomly select $M$ samples from set $\Omega$ and express them as $x^{(1)},\dots,x^{(M)}$. Their costs are supposed to follow an identical Gaussian distribution $p(F) = \mathcal{N}[\mu, \sigma^2]$ with mean $\mu$ and variance $\sigma^2$. The minimum among these $M$ samples is denoted by $F_M$. The cumulative distribution function of $F$ is defined as $\varphi(F) \equiv \int_{-\infty}^F p(F') dF'$. Moreover, the cumulative distribution function of $F_M$ is given by $\varphi_M(F_M) \equiv \mathrm{Prob}[F_M \leq F(x^{(1)}),\dots,F(x^{(M)})] = 1 - \mathrm{Prob}[F_M > F(x^{(1)}),\dots,F(x^{(M)})]$. Since sampling is independent, it becomes $\varphi_M(F_M) = 1 - \prod_{1 \leq t \leq M}\mathrm{Prob}[F_M > F(x^{(t)})] = 1 - (1 - \varphi(F_M))^M$. As the derivative of $\varphi_M(F_M)$, the probability density of $F_M$ is given by
\begin{equation}
\label{probability_density}
p_M(F_M) = M (1 - \varphi(F_M))^{M-1} p(F_M).
\end{equation}
The mode of $p_M(F_M)$ satisfies
\begin{equation}
\label{first_derivative}
\frac{dp_M(F_M)}{dF_M} = -M(M-1)(1 - \varphi(F))^{M-2}p(F_M)^2 + M (1 - \varphi(F))^{M-1} \left.\frac{dp(F)}{dF}\right|_{F=F_M} = 0.
\end{equation}
When $F_M$ is sufficiently smaller than $\mu$, since $M \gg 1$ and $\varphi(F_M) \ll 1$, \eqref{first_derivative} is approximated as
\begin{equation}
\label{mode}
M p(F_M) \approx -\frac{F_M - \mu}{\sigma^2}.
\end{equation}
Here, we used $dp(F)/dF = -(F - \mu)/\sigma^2 p(F)$. Thus, we find the following theorem:

\paragraph{\em Theorem 1: required iterations for random search to obtain $F_M$.}
{\em When $F_M \ll \mu$, the required iterations $M$ to attain $F_M$, with a $1/2$ probability, by randomly selecting $F(x)$ from $p(F) = \mathcal{N}[\mu,\sigma^2]$ are}
\begin{equation}
M \approx \sqrt{\frac{2\pi(F_M - \mu)^2}{\sigma^2}} \exp\left(\frac{(F_M - \mu)^2}{2\sigma^2}\right).
\end{equation}

Furthermore, taking the logarithm of \eqref{mode} yields
\begin{equation}
\label{mode_log}
\frac{(F_M - \mu)^2}{2\sigma^2} + \log(-F_M + \mu) = \frac{1}{2}\log\frac{M^2\sigma^2}{2\pi}.
\end{equation}
As the leading order, it approximately holds that $-F_M + \mu \approx \sqrt{\sigma^2\log{M^2\sigma^2}}$. By substituting it into $\log(-F_M + \mu)$ of \eqref{mode_log}, we obtain the mode of $p_M(F_M)$ as
\begin{equation}
\label{mode2}
F_M = \mu - \sigma \sqrt{\log\frac{M^2\sigma^2}{2\pi} - \log(\sigma^2\log{M^2\sigma^2})} = \mu - \sigma \sqrt{\log\frac{M^2}{2\pi\log{M^2\sigma^2}}}.
\end{equation}
Then, we approximate $p_M(F_M)$ as a Gaussian distribution, $p_M(F_M) = \mathcal{N}[\mu_M, \sigma_M^2]$ (Laplace approximation). The mean, $\mu_M$, matches the mode of $p_M(F_M)$. Moreover, from the second derivative of $p_M(F_M)$, the variance of $p_M(F_M)$ is solved as $\sigma_M^2 = \sigma^2 \left( \log(M^2 / 2\pi\log{M^2\sigma^2}) - 1 \right)^{-1}$. See Supplementary Methods S1 for the derivation. Therefore, we find the following theorem:

\paragraph{\em Theorem 2: distribution of minimum among randomly selected M samples.}
{\em When $M$ samples are independently and randomly generated from an identical distribution $p(F) = \mathcal{N}[\mu, \sigma^2]$, the minimum among these samples, $F_M$, approximately follows}
\begin{align}
\label{theorem2}
p_M(F_M) = \mathcal{N}\left[ \mu - \sigma \sqrt{\log\frac{M^2}{2\pi\log{M^2\sigma^2}}}, \;\;\; \sigma^2 \left( \log\frac{M^2}{2\pi\log{M^2\sigma^2}} - 1 \right)^{-1} \right] \nonumber \\
\approx \mathcal{N}\left[ \mu - \sigma \sqrt{2\log M}, \;\;\; \frac{\sigma^2}{2\log{M}} \right]. \hspace{30mm}
\end{align}

From Theorem 2, we can obtain the probability of an approximate global minimum of the system. When $p(F) = \mathcal{N}[0,1]$ and $M = 2^N$, the distribution of an approximate global minimum $F_{2^N}$ converges to a delta function
\begin{equation}
\label{global_minimum}
p_{2^N}(F_{2^N}) \approx \delta\left( F_{2^N} + \sqrt{2N\log{2}} \right).
\end{equation}
When $F(x)$ involves only lower-order products, $F_{2^N}$ can be slightly close to zero due to a correlation between costs of neighboring states; however, the difference is expected to be small. In what follows, we will mathematically analyze the the computational cost to identify a state providing the above-mentioned approximate global minimum using the selection and crossover algorithm.


\section{Theory on the computational cost of the selection and crossover algorithm}

Here, it is demonstrated that the selection and crossover algorithm accelerates the global search. In this work, a crossover (or a mixture) of $n$ favorable states, $x^{(1)},\dots,x^{(n)}$, is expressed as an operation that randomly generates offspring $z = (z_1,\dots,z_N) \in Z$ by following the probability distribution
\begin{equation}
\mathrm{Prob}[z_i = 1] = \rho_i \equiv \frac{x^{(1)}_i + \cdots + x^{(n)}_i}{2n} + \frac{1}{2}.
\end{equation}
The mean and variance of $z_i$ are $\E_Z[z_i] = 2\rho_i - 1$ and $\Var_Z[z_i] = 1 - \E_Z[z_i]^2 = 4\rho_i (1-\rho_i)$; thus, we can see that $0 \leq \Var_Z[z_i] \leq 1$ and the variance is maximized when and only when $\rho_i = 1/2$. The case with $n = 2$ provides the selection and crossover algorithm, while a large $n$ limit is known as the mean-field approximation, where elements of a state are assumed to be independent from each other. When $n = 2$, it holds that $\rho_i = 1/2$ when $x^{(1)}_i = -x^{(2)}_i$ and $\rho_i \in \{0,1\}$ otherwise. In contrast, when $n$ is larger than 2, $0 < \rho_i < 1$ and $\rho_i \neq 1/2$ hold for a generic cost function; thus, the variance is not maximized in this condition. Below, it is mathematically expressed that $n = 2$ is the best to maximize the variance (i.e., variety) of offspring's distribution; therefore, the selection and crossover algorithm can accelerate the global search.


\subsection{Distribution of a crossover of two favorable states}

Here, we consider the case with $n = 2$. We randomly generate $M$ samples from $X$ and select the minimum, referred to as $x$; then, we randomly generate another $M$ samples and select the minimum, referred to as $y$ (Fig. 1B). The computational cost for this operation is $2M$. From Theorem 2, the results of $F(x)$ and $F(y)$ independently follow an identical Gaussian distribution \eqref{theorem2}.

We define a schema $x^*$ as a set of common elements between $x$ and $y$ (i.e., parents) and $Z$ as a set of $x$ that involves $x^*$. The expectation of $F(z)$ over $z \in Z$ (i.e., offspring) is likely to be lower than that over $x \in X$. We define the number of different elements by $d \equiv |x-y|^2/4$ using the Euclidean distance. Without loss of generality, we can suppose that $x_i$ and $y_i$ are different for $1 \leq i \leq d$ (i.e., $x_i = -y_i$), while $x_i$ and $y_i$ are the same for $d + 1 \leq i \leq N$ (i.e., $x_i=y_i$). Thus, $z_1, z_2, \dots, z_d$ are randomly sampled from $\{-1,1\}^d$ while $(z_{d+1}, \dots, z_N) = (x_{d+1}, \dots, x_N)$ is fixed. This can be viewed as a problem with a $d$-dimensional state space.

It is possible to decompose $F(z)$ by the number of random variables $(z_1, z_2, \dots, z_d)$ involved in a term. We define terms involving the $\alpha$-th order products of $z_1, z_2, \dots, z_d$ as follows:
\begin{equation}
\begin{cases}
\; C_0 & \equiv \displaystyle \sum_{d+1 \leq i \leq N} \!\!\!\! a_i x_i + \sum_{d+1 \leq i < j \leq N} \!\!\!\!\!\! a_{ij} x_i x_j + \sum_{d+1 \leq i < j < k \leq N} \!\!\!\!\!\!\!\! a_{ijk} x_i x_j x_k + \cdots \\
\; C_1(z) & \equiv \displaystyle \sum_{1 \leq i \leq d} a_i z_i + \sum_{\substack{1 \leq i \leq d \\ d+1 \leq j \leq N}} \!\!\!\! a_{ij} z_i x_j + \sum_{\substack{1 \leq i \leq d \\ d+1 \leq j < k \leq N}} \!\!\!\!\!\! a_{ijk} z_i x_j x_k + \cdots \\
\hspace{10mm} & \; \vdots
\end{cases}  
\end{equation}
In general, these terms are expressed as
\begin{equation}
C_\alpha(z) \equiv \sum_{1 \leq i_1 < \cdots < i_\alpha \leq d} \!\!\!\!\!\! b_{i_1 \dots i_\alpha} z_{i_1} \dots z_{i_\alpha}
\end{equation}
for $0 \leq \alpha \leq d$, where coefficients are given as
\begin{equation}
b_{i_1 \dots i_\alpha} \equiv a_{i_1 \dots i_\alpha} + \sum_{d+1 \leq j_1 \leq N} \!\!\!\! a_{i_1 \dots i_\alpha j_1} x_{j_1} + \sum_{d+1 \leq j_1 < j_2 \leq N} \!\!\!\!\!\! a_{i_1 \dots i_\alpha j_1 j_2} x_{j_1} x_{j_2} + \cdots.
\end{equation}


The mean of $b_{i_1 \dots i_\alpha}$ becomes $\E_{A,X^*}[b_{i_1 \dots i_\alpha}] = 0$ since $a_{i_1 \dots i_\alpha}, a_{i_1 \dots i_\alpha j_1}, \dots$ follow zero-mean distributions, where $X^*$ is a set of $x^*$, and the variance of $b_{i_1 \dots i_\alpha}$ is calculated as
\begin{align}
\E_A\Big[ \Var_{X^*}\left[b_{i_1 \dots i_\alpha}\right] \Big] = \E_A[a_{i_1 \dots i_\alpha}^2] + \!\! \sum_{d+1 \leq j_1 \leq N} \!\!\!\!\!\! \E_A[a_{i_1 \dots i_\alpha j_1}^2]\E_{X^*}[x_{j_1}^2] + \cdots \hspace{11mm} \nonumber \\
= \sigma_\alpha^2 + \binom{N-d}{1}\sigma_{\alpha+1}^2 + \binom{N-d}{2}\sigma_{\alpha+2}^2 + \cdots = \sum_{0 \leq \beta \leq N-d} \binom{N-d}{\beta}\sigma_{\alpha+\beta}^2,
\end{align}
since $a_{i_1 \dots i_\alpha}, a_{i_1 \dots i_\alpha j_1}, \dots$ are independent from each other. Thus, we find that $b_{i_1 \dots i_\alpha}$ follows the following Gaussian distribution:
\begin{equation}
p(b_{i_1 \dots i_\alpha}) = \mathcal{N}\left[ 0, \sum_{0 \leq \beta \leq N-d} \binom{N-d}{\beta}\sigma_{\alpha+\beta}^2 \right].
\end{equation}
Since $\E_Z[C_\alpha(z)] = 0$ and $\Var_Z[C_\alpha(z)] = \E_Z[C_\alpha(z)^2] = \sum_{1 \leq i_1 < \cdots < i_\alpha \leq d} b_{i_1 \dots i_\alpha}^2$ for $1 \leq \alpha \leq d$, it holds that
\begin{equation}
\label{C(z)_variance}
\E_{A,X^*}\Big[ \E_Z[C_\alpha(z)^2] \Big] = \binom{d}{\alpha} \sum_{0 \leq \beta \leq N-d} \binom{N-d}{\beta}\sigma_{\alpha+\beta}^2.
\end{equation}


Thus, $F(z)$ is expressed as $F(z) = C_0 + C_1(z) + C_2(z) + \cdots + C_d(z) = \sum_{0 \leq \alpha \leq d} C_\alpha(z)$. The distribution of $F(z)$ over $Z$ has the mean $\E_Z[F(z)] = C_0$ and the variance $\Var_Z[F(z)] = \E_Z[F(z)^2] - C_0^2 = \E_Z[(C_1(z) + \cdots + C_d(z))^2]$. If we assume that $C_1(z), \dots, C_d(z)$ are independent from each other, from \eqref{C(z)_variance}, we obtain
\begin{equation}
\E_{A,X^*}\!\Big[ \Var_Z[F(z)] \Big] = \E_{A,X^*}\!\Big[ \E_Z[C_1(z)^2 + \cdots + C_d(z)^2] \Big] = \sum_{1 \leq \alpha \leq d} \!\! \binom{d}{\alpha} \!\! \sum_{0 \leq \beta \leq N-d} \!\! \binom{N-d}{\beta}\sigma_{\alpha+\beta}^2.
\end{equation}
Using the Vandermonde's identity, we find
\begin{align}
\E_{A,X^*}\Big[ \Var_Z[F(z)] \Big] = \sum_{0 \leq \alpha \leq d} \binom{d}{\alpha} \sum_{0 \leq \gamma \leq N} \binom{N-d}{\gamma - \alpha} \sigma_{\gamma}^2 - \sum_{0 \leq \beta \leq N-d}\binom{N-d}{\beta}\sigma_\beta^2 \nonumber \\
= \sum_{0 \leq \gamma \leq N} \binom{N}{\gamma} \sigma_{\gamma}^2 - \sum_{0 \leq \beta \leq N-d}\binom{N-d}{\beta}\sigma_\beta^2 = 1 - \sum_{0 \leq \beta \leq N-d}\binom{N-d}{\beta}\sigma_\beta^2,
\end{align}
where $\gamma \equiv \alpha + \beta$. The last equality holds from \eqref{variance}.


Next, we estimate the expectation of $C_0$ over various $a_{i_1 \dots i_\alpha} \in A$. From $F(x) = \sum_{0 \leq \alpha \leq d} C_\alpha(x)$, $F(y) = \sum_{0 \leq \alpha \leq d} C_\alpha(y)$, and $(y_1, \dots, y_d) = -(x_1, \dots, x_d)$, it holds that
\begin{equation}
\begin{cases}
\; \displaystyle \frac{F(x) + F(y)}{2} = C_0(x) + C_2(x) + C_4(x) + \cdots = -\sqrt{\log\frac{M^2}{2\pi\log{M^2}}} \\
\; \displaystyle \frac{F(x) - F(y)}{2} = C_1(x) + C_3(x) + C_5(x) + \cdots = 0
\end{cases}
\end{equation}
Here, one can view that, in both $F(x)$ and $F(y)$, $C_0(x), C_2(x), C_4(x), \dots$ have been optimized to minimize the sum of them. Whereas, $C_1(x), C_3(x), C_5(x), \dots$ do not contribute to the optimization since the sum of them is zero, which implies that they are negligible when considering the expectation of a cost over various $a_{i_1 \dots i_\alpha} \in A$. Thus, we will drop them from evaluation. It is further assumed that the optimization using the random search is unbiased in the sense that the expectations of $-C_0(x), -C_2(x), -C_4(x), \dots$ over various $a_{i_1 \dots i_\alpha} \in A$ are simply proportional to their magnitudes. Expressed differently, we assume
\begin{equation}
\frac{\E_A[C_0(x)]}{\Var_{A,X}[C_0(v)]} = \frac{\E_A[C_2(x)]}{\Var_{A,X}[C_2(v)]} = \frac{\E_A[C_4(x)]}{\Var_{A,X}[C_4(v)]} = \cdots,
\end{equation}
where $x$ indicates the solution of the above-mentioned random search given a particular cost function, while $v \in X$ expresses a state randomly selected from $X$. This assumption can be viewed as a flat prior belief about cost functions from a Bayesian perspective.

Under these assumptions, it holds that $\E_A[C_0(x)] = \E_A[C_0(x) + C_2(x) + \cdots] \cdot \Var_{A,X}[C_0(v)] / \Var_{A,X}[C_0(v) + C_2(v) + \cdots]$. For a large $N$, it is approximated that
\begin{equation}
\Var_{A,X}[C_0(v) + C_2(v) + \cdots] = \sum_{\alpha=0,2,4,\dots} \sum_{1 \leq \beta \leq N-d} \binom{N-d}{\beta}\sigma_{\alpha+\beta}^2 \approx \frac{1}{2} \sum_{1 \leq \gamma \leq N} \binom{N}{\gamma} \sigma_{\gamma}^2 = \frac{1}{2}
\end{equation}
from the Vandermonde's identity and \eqref{variance}. Moreover, for a large $N$, $d$ converges to $d = N/2$. Therefore, the expectation of $C_0$ is estimated as
\begin{equation}
\E_A[C_0] = -2 \left( \sum_{1 \leq \beta \leq N/2} \binom{N/2}{\beta}\sigma_\beta^2 \right) \sqrt{\log\frac{M^2}{2\pi\log{M^2}}}.
\end{equation}

Thus, under the Laplace assumption, we obtain the following theorem:

\paragraph{\em Theorem 3: distribution of crossovers.}
{\em Under the above-mentioned assumptions, $z \in Z$ provides $F(z)$ that follows}
\begin{equation}
\label{theorem3_1}
p(F) = \mathcal{N}\left[ -2 \eta \sqrt{\log\frac{M^2}{2\pi\log{M^2}}}, \;\;\;\; 1 - \eta \right],
\end{equation}
{\em where $\eta \equiv \sum_{1 \leq \beta \leq N/2} \binom{N/2}{\beta}\sigma_\beta^2$. Therefore, when $M$ samples are independently and randomly generated from offspring's distribution, from Theorem 2, the minimum among these samples, $F_M$, approximately follows}
\begin{equation}
\label{theorem3_2}
p_M(F_M) = \mathcal{N}\left[ -\left(2\eta + \sqrt{1-\eta}\right) \sqrt{\log\frac{M^2}{2\pi\log{M^2}}}, \;\;\;\; (1 - \eta) \left( \log\frac{M^2}{2\pi\log{M^2}} - 1 \right)^{-1} \right].
\end{equation}

Hence, by solving $-(2\eta + \sqrt{1-\eta})\sqrt{2\log{M}} = -\sqrt{2N\log{2}}$, we find that

\paragraph{\em Theorem 4: computational cost for the selection and crossover algorithm.}
{\em The computational cost (total iterations) $M$ to attain an approximate global minimum using the selection and crossover algorithm are}
\begin{equation}
M \approx 3 \cdot \lambda^N, \quad \mathrm{where} \quad \lambda \equiv 2^{\frac{1}{(2\eta + \sqrt{1-\eta})^2}} < 2.
\end{equation}

In summary, we have derived the distribution of offspring generated by a crossover of two favorable states, which is characterized by negative mean and asymptotically unit variance. Consequently, the selection and crossover algorithm reduces the computational cost for the global search to $\mathcal{O}(\lambda^N)$.



\subsection{Distribution of a mixture of many favorable states}
For comparison, when $n$ is large, the mean and variance of offspring's distribution are calculated as follows. Because $0 < \rho_i < 1$ for a generic cost function, from \eqref{cost_function}, we obtain the mean $\E_Q[F(z)] = R \equiv \sum_{1 \leq \alpha \leq N} \left( \sum_{1 \leq i_1 < \cdots < i_\alpha \leq N} a_{i_1 \dots i_\alpha} \rho_{i_1} \cdots \rho_{i_\alpha} \right) \neq 0$, where the probability of $z$, $Q \equiv Q(z) = \prod_{1 \leq i \leq N}\mathrm{Prob}(z_i)$, is characterized by $\rho_1, \dots, \rho_N$. Moreover, we obtain the variance $\Var_Q[F(z)] = 1 - \E_Q[F(z)]^2 = 1 - R^2 < 1$. From \eqref{mode2}, the expectation of the minimum among randomly selected $M$ samples from the distribution is $F_{M} = R - \sqrt{(1 - R^2) \cdot 2\log{M}} \geq -\sqrt{1 + 2\log{M}}$, where $F_{M} = -\sqrt{1 + 2\log{M}}$ holds when and only when $R = -1/\sqrt{1+2\log{M}}$. Therefore, a mixture of many favorable states is, at most, the same performance as the random search for a large $M$, and thus ineffective for the global search.


\section{Simulation and results}

\begin{figure}[t]
  \centering
  \includegraphics[width=0.98\linewidth]{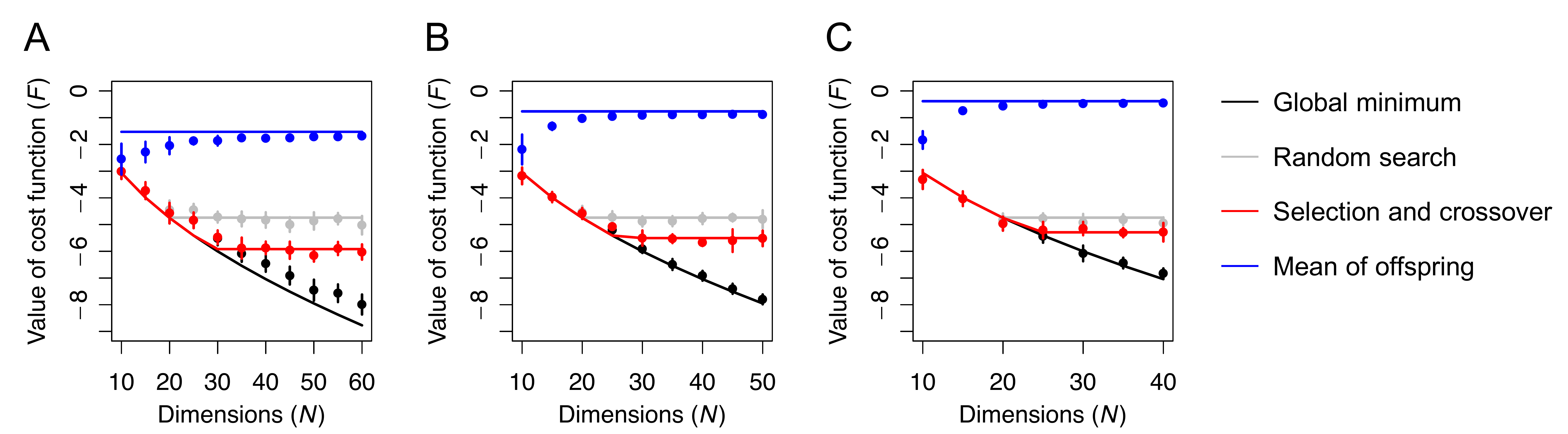}
  \caption{Simulation results of search algorithms. We consider optimization of the $K$-th order cost function, where the variances of coefficients satisfy $\sigma_1^2 = \cdots = \sigma_K^2$ and $\sigma_{K+1}^2 = \cdots = \sigma_N^2 = 0$. (A), (B), and (C) illustrate optimization of the second-, third-, and fourth-order cost function, respectively. Global minimum (black) is determined by sequentially searching all possible states for $N \leq 20$, or approximated by the best solution of gradient descent searches with iterative random initialization ($10^3$ times) otherwise. Random search (gray) is conducted by randomly sampling states for $10^6$ times. Selection and crossover algorithm (red) determines parents as states providing the minimum cost among two different $10^3$ samples, and generates offspring as a crossover of parents' states, followed by the selection of offspring providing the minimum cost among $10^3$ samples. This process is repeated $333$ times with different parents and the best solution is selected, to ensure the total computational cost equivalent to the random search. The mean of offspring's distribution (blue) takes a negative value, which converges to zero as $K$ increases. Circles and bars indicate the mean and standard deviation obtained by 10 different cost functions. Lines are theoretical predictions.}
\end{figure}

The accuracy of the above-mentioned theoretical predictions is examined by numerical simulations (codes are appended as Supplementary Source Codes). Here, we consider a class of the $K$-th order cost functions comprising up to the $K$-th order products of elements of $x$. Figure 2 illustrates the relationships between the dimensions of state space $N$ and the value of the global minimum, the solutions of random search and selection and crossover algorithm, and the mean of offspring's distribution obtained by a crossover of two favorable states, when the second-, third- and fourth-order cost function is considered (A-C). First, the global minimum becomes large negative in the order of $\sqrt{N}$ as predicted by \eqref{global_minimum}, although the global minimum for $K = 2$ is slightly larger than the prediction due to the considerable correlation between neighboring states. The random search provides a solution with a cost determined only by the number of iterations ($M$), as shown in \eqref{theorem2}, regardless the order $K$ and dimensions $N$, unless it finds the global minimum.

Crucially, the selection and crossover algorithm finds a solution with a cost $F$ lower than that of the random search, while their computational costs are the same. The outcome of using the selection and crossover algorithm is characterized by the gain $g \equiv (2\eta + \sqrt{1-\eta})^2 > 1$, where the search efficacy of the selection and crossover algorithm with the computational cost $M$ is equivalent, in the order, to that of iterating random searches $M^g$ times, providing a higher chance to find the global minimum. The gain becomes close to 2 when $K = 2$, while it converges to 1 when $K$ is large due to an increasing complexity of a cost function. The mean of the offspring's distribution is proportional to $-\eta$, where $\eta$ converges to $2^{-K}$ for the $K$-th order cost functions when $N$ is large. These properties are as predicted by Theorem 3: see \eqref{theorem3_1} and \eqref{theorem3_2}.

In contrast, when we generate offspring following a mixture of several favorable states, as an mean-field approximation of their probability distribution, the best solution among the offspring's distribution is much worse (i.e., higher) than that obtained by a crossover of two favorable states, while both methods iterate the selection of parents and offspring with the same number $M$ (Fig. 3A). The difference between a crossover of two states and a mixture of several states can be explained by the variances of their offspring (Fig. 3B). The variance of the distribution created by selection and crossover is large and converges to 1 as $K$ increases, indicating that sampling from the offspring's distribution has sufficient variety. Whereas, the near-zero variance of the distribution created by the mean-field approximation indicates a near-zero variety of offspring. Therefore, only the selection and crossover algorithm can retain the variety of offspring and use it to enhance the search efficacy.



\begin{figure}[t]
  \centering
  \includegraphics[width=0.98\linewidth]{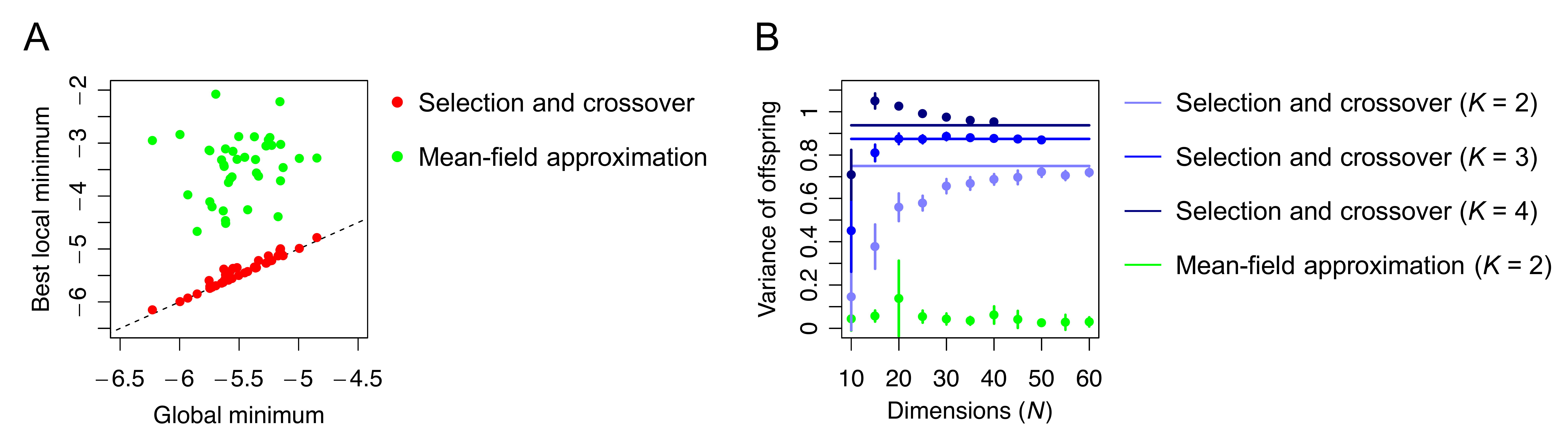}
  \caption{Comparison between selection and crossover algorithm and mean-field approximation. (A) Optimization of the second-order cost function with $N = 30$. In most cases, selection and crossover algorithm (red), with the same computational cost as in Fig. 2, finds the global minimum. In contrast, a mixture of four favorable states (green), which approximates a probability distribution of favorable states under an mean-field approximation, does not efficiently find the global minimum, despite the same number of iterations as the selection and crossover algorithm. (B) Variance of offsprings obtained using selection and crossover algorithm and mean-field approximation. Circles, bars, and lines are the same meanings as in Fig. 2.}
\end{figure}


\section{Discussion and conclusion}

An optimization problem comprising a cost function and discrete states can be solved in principle; however, the computational cost may be large, in the worst case increasing to the $2^N$ order of the dimensions of states $N$. Consequently, in practice, the optimization problem cannot be solved. This work analytically and numerically demonstrates the computational cost of the selection and crossover algorithm for determining an approximate global minimum of a class of cost functions characterized by the variance of coefficients. The use of the gradient descent search should accelerate the global search. In future work, the computational cost for an approach combining the gradient descent search and the selection and crossover algorithm will be addressed.

While the selection and crossover algorithm has been popularized as a genetic algorithm \cite{Mitchell1998, Sivanandam_Deepa2007, Weise2009}, it should be emphasized that it can be a computational model for insightful search in the brain \cite{Jung-Beeman2004, Bowden2005}. Creating a new state with a crossover of two existing states may be more relevant than randomly coming up with a new state, as a function of insight, in the sense that only the crossover can enhance the search efficacy. The present work is potentially interesting because optimizations in neurobiology are in most cases modeled as local searches; thus, the selection and crossover algorithm may provide an explanation about optimizations beyond local searches which the brain may perform.

In summary, it is shown that when a random search needs iterations in the order of $2^N$ to determine an approximate global minimum, the selection and crossover algorithm can reduce the computational cost to the order of $\lambda^N$ with $\lambda$ less than 2. Unlike a mixture of several favorable states, a crossover of two favorable states can provide a distribution of offspring with negative mean and variance that converges to 1. This highlights an advantage of a crossover of two favorable states, as opposed to a mixture of many favorable states, in efficiently determining an approximate global minimum.



\subsubsection*{Acknowledgments}
This work was supported by RIKEN Center for Brain Science.



\newpage

\section*{Supplementary Information}
\section*{Computational cost for determining an approximate global minimum using the selection and crossover algorithm}

\vspace{5mm}

\section*{Supplementary Tables}
\begin{table}[h]
  \text{Table S1: Glossary of expressions}
  \centering
  \begin{tabular}{ll}
    \toprule
    Expression					& Description							\\
    \midrule
    $N$						& Input size (dimension)					\\
    $x = (x_1,x_2,\dots,x_N)$		& State								\\
    $X \equiv \{-1,1\}^N$			& Set of states							\\
    $\Omega \subseteq X$			& Arbitrary set of states					\\
    $F(x)$						& Cost function							\\
    $a_{i_1 \dots i_\alpha} \in A$		& Coefficient							\\
    $A$						& Distribution of coefficients				\\
    $M$						& Computational cost (number of iterations)	\\
    \midrule
    $z = (z_1,z_2,\dots,z_N) \in Z$	& Offspring state						\\
    $Z \subseteq X$				& Set of offspring states					\\
    $x^* \in X^*$					& Schema, i.e., common elements between parents	\\
    $X^*$						& Set of schemata						\\
    $d \equiv |x - y|^2/4$			& Number of different elements				\\
    $C_\alpha$					& Sum of the $\alpha$-th order products of $z_1, \dots, z_d$	\\
    $\eta$						& Magnitude of $C_0$					\\
    $g \equiv (2\eta+\sqrt{1-\eta})^2$	& Gain								\\
    \midrule
    $\E_X[F(x)]$					& Expectation of $F(x)$ over $x$ randomly sampled from $X$		\\
    $\Var_X[F(x)]$				& Variance of $F(x)$ over $x$ randomly sampled from $X$		\\
    $\E_A[\Var_X[F(x)]]$			& Expectation of $\Var_X[F(x)]$ over various $a_{i_1 \dots i_\alpha} \in A$	\\
    $\E_{X^*}[b_{i_1 \dots i_\alpha}]$	& Expectation of $b_{i_1 \dots i_\alpha}$ over $x^* \in X^*$		\\
    $\Var_{X^*}[b_{i_1 \dots i_\alpha}]$	& Variance of $b_{i_1 \dots i_\alpha}$ over $x^* \in X^*$		\\
    $\E_{A,X^*}[b_{i_1 \dots i_\alpha}] \equiv \E_A[\E_{X^*}[b_{i_1 \dots i_\alpha}]]$	& Expectation of $\E_{X^*}[b_{i_1 \dots i_\alpha}]$ over $a_{i_1 \dots i_\alpha} \in A$	\\
    $\E_A[\Var_{X^*}[b_{i_1 \dots i_\alpha}]]$	& Expectation of $\Var_{X^*}[b_{i_1 \dots i_\alpha}]$ over $a_{i_1 \dots i_\alpha} \in A$	\\
    \bottomrule
  \end{tabular}
\end{table}

\section*{Supplementary Methods}
\subsection*{S1. Derivation of variance of $p_M(F_M)$}
Since the first derivative of $p_M(F_M)$ is expressed as \eqref{probability_density}, the second derivative is obtained as
\begin{align}
& \frac{d^2p_M(F_M)}{dF_M^2} = M(M-1)(M-2)(1-\varphi(F_M))^{M-3} p(F_M)^3 \nonumber \\
& - 3M(M-1)(1-\varphi(F_M))^{M-2} p(F_M)\left.\frac{dp(F)}{dF}\right|_{F=F_M} + M(1-\varphi(F_M))^{M-1}\left.\frac{d^2p(F)}{dF^2}\right|_{F=F_M} \nonumber \\
& = \Bigg\{ (M-1)(M-2)(1-\varphi(F_M))^{-2} p(F_M)^2 - 3(M-1)(1-\varphi(F_M))^{-1} \left.\frac{dp(F)}{dF}\right|_{F=F_M} \nonumber \\
& + \frac{1}{p(F_M)} \left.\frac{d^2p(F)}{dF^2}\right|_{F=F_M} \Bigg\} p_M(F_M).
\end{align}
Since $M \gg 1$ and $\varphi(F_M) \ll 1$, it holds that
\begin{align}
\frac{1}{p_M(F_M)}\frac{d^2p_M(F_M)}{dF_M^2} \approx M^2 p(F_M)^2 - 3M \frac{-(F_M - \mu)}{\sigma^2} p(F_M) + \frac{1}{\sigma^2} + \frac{(F_M - \mu)^2}{\sigma^4} \nonumber \\
\approx \frac{(F_M - \mu)^2}{\sigma^4} - 3\frac{(F_M - \mu)^2}{\sigma^4} + \frac{1}{\sigma^2} + \frac{(F_M - \mu)^2}{\sigma^4} = -\frac{(F_M - \mu)^2}{\sigma^4} + \frac{1}{\sigma^2}.
\end{align}
Here, in the second line, we used \eqref{mode}. Moreover, when $p_M(F_M) = \mathcal{N}[\mu_M, \sigma_M^2]$, the second derivative is given by $d^2p_M(F_M)/dF_M^2 = -1/\sigma_M^2 \cdot p_M(F_M) + (F_M - \mu_M)^2/\sigma_M^4 \cdot p_M(F_M)$. Thus, when $F_M = \mu_M$, from \eqref{mode2}, it is obtained that
\begin{equation}
\frac{1}{\sigma_M^2} = \frac{1}{\sigma^2} \left( \frac{(F_M - \mu)^2}{\sigma^2} - 1 \right) = \frac{1}{\sigma^2} \left( \log\frac{M^2}{2\pi\log{M^2\sigma^2}} - 1 \right).
\end{equation}


\end{document}